\documentclass{aa}  
\usepackage{graphicx}
\usepackage{txfonts}
\usepackage{footnote}
\usepackage[colorlinks=true,linkcolor=blue,citecolor=blue,filecolor=blue,urlcolor=blue]{hyperref}
\usepackage{subcaption} 
\usepackage{amssymb} 
\usepackage{ulem} 
\usepackage[dvipsnames]{xcolor} 
\usepackage{siunitx}
\usepackage{ gensymb } 
\usepackage[flushleft]{threeparttable} 
\usepackage{soul} 
\newcommand{\orcid}[1]{\href{https://orcid.org/#1}{\textcolor[HTML]{A6CE39}{\aiOrcid}}}
\DeclareUnicodeCharacter{2212}{-}
\usepackage{multirow}
\usepackage{multicol}
\usepackage{url}

\usepackage{fancyhdr}
\usepackage{natbib}
\bibpunct{(}{)}{;}{a}{}{,}

\newcommand{\qhired}{$\chi^{2}_{\nu}$}

\newcommand{\Msun}{M$_\mathrm{\odot}$}
\newcommand{\Rsun}{R$_\mathrm{\odot}$}
\newcommand{\ergs}{erg~s$^{-1}$}
\newcommand{\flux}{\mathrm{erg~cm}^{-2}~\mathrm{s}^{-1}}

\defcitealias{CorralSantana2013}{CS13}
\defcitealias{JimenezIbarra2019b}{JI19c}
\defcitealias{JimenezIbarra2019c}{JI19b}

\begin{document} 

  \title{The omnipresent flux-dependent optical dips of the black hole transient Swift~J1357.2-0933}

  \titlerunning{The omnipresent flux-dependent optical dips of the black hole transient Swift~J1357.2-0933}
  
    \author{G. Panizo-Espinar \inst{1}\textsuperscript{,}\inst{2}\fnmsep\thanks{E-mail: guayente.panizo@gmail.es}
       \and T.~Muñoz-Darias\inst{1}\textsuperscript{,}\inst{2}
       \and M.~Armas~Padilla\inst{1}\textsuperscript{,}\inst{2}
       \and F.~Jiménez-Ibarra\inst{3}
       \and D.~Mata~Sánchez\inst{1}\textsuperscript{,}\inst{2}
       \and I.~V.~Yanes-Rizo \inst{1}\textsuperscript{,}\inst{2}
       \and K.~Alabarta\inst{4}\textsuperscript{,}\inst{5}
       \and M.~C.~Baglio\inst{6}
       \and E.~Caruso\inst{7}
       \and J.~Casares\inst{1}\textsuperscript{,}\inst{2}
       \and J.~M.~Corral-Santana\inst{8}
       \and F.~Lewis\inst{9}\textsuperscript{,}\inst{10}
       \and D.~M.~Russell\inst{4}\textsuperscript{,}\inst{5}
       \and P.~Saikia\inst{4}\textsuperscript{,}\inst{5}
       \and J.~Sánchez-Sierras\inst{1}\textsuperscript{,}\inst{2}
       \and T.~Shahbaz\inst{1}\textsuperscript{,}\inst{2}
       \and M.~A.~P.~Torres\inst{1}\textsuperscript{,}\inst{2}
       \and F.~Vincentelli\inst{1}\textsuperscript{,}\inst{2}}
\authorrunning{Panizo-Espinar et al.} 
   \institute{Instituto de Astrof\'isica de Canarias (IAC), V\'ia Láctea, La Laguna, E-38205, Santa Cruz de Tenerife, Spain
    \and 
    Departamento de Astrof\'isica, Universidad de La Laguna, E-38206 Santa Cruz de Tenerife, Spain 
    \and    
    School of Physics and Astronomy, Monash University, Clayton, Victoria 3800, Australia
    \and Center for Astrophysics and Space Science (CASS), New York University Abu Dhabi, PO Box 129188, Abu Dhabi, UAE
    \and New York University Abu Dhabi, PO Box 129188, Abu Dhabi, United Arab Emirates
    \and INAF, Osservatorio Astronomico di Brera, Via E. Bianchi 46, I-23807 Merate (LC), Italy
    \and Anton Pannekoek Institute for Astronomy, University of Amsterdam, Science Park 904, NL-1098 XH Amsterdam, the Netherlands
   \and European Southern Observatory, Alonso de C\'ordova 3107, Vitacura, Casilla 19001, Santiago de Chile, Chile
    \and Faulkes Telescope Project, School of Physics and Astronomy, Cardiff University, The Parade, Cardiff, CF24 3AA, Wales, UK 
    \and Astrophysics Research Institute, Liverpool John Moores University, 146 Brownlow Hill, Liverpool L3 5RF, UK
    }

  \abstract
  {Swift J1357.2-0933 is a black hole transient of particular interest due to the optical, recurrent dips found during its first two  outbursts (in 2011 and 2017), with no obvious X-ray equivalent. We present fast optical photometry during its two most recent outbursts, in 2019 and 2021. Our observations reveal that the optical dips were present in every observed outburst of the source, although they were shallower and showed longer recurrence periods in the two most recent and fainter  events. We perform a global study of the dips properties in the four outbursts, and find that they do not follow a common temporal evolution. In addition, we discover a correlation with the X-ray and optical fluxes, with dips being more profound and showing shorter recurrence periods for brighter stages. This trend seems to extend even to the faintest, quiescent states of the source. Finally, we discuss these results in the context of the possible connection between optical dips and outflows found in previous works. }
  

   \maketitle
\section{Introduction}\label{cap.introduction}

Low-mass X-ray binaries (LMXBs) are stellar systems comprised of either a black hole (BH) or a neutron star and a $\lesssim$~1~\Msun \ mass donor (e.g., \citealt{Casares2017, Bahramian2023}). The companion star transfers gas via Roche-lobe  overflow, originating an accretion disc where matter can reach temperatures high enough ($\sim$10$^{7}$ K) to emit in X-rays. Most of these systems, the so-called X-ray transients, spend the majority of their lives in a quiescent, dim state, showing only occasional episodes of enhanced accretion. During these outbursts,  usually lasting from a few weeks to months, their X-ray luminosities can increase from $\sim$10$^{31-34}$ \ergs \ to $\sim$10$^{36-39}$ \ergs \  and a complex phenomenology of different accretion states and associated outflows is observed (e.g., \citealt{Wijnands2006, Belloni2011, Fender2016}). 

Swift~J1357.2-0933 (J1357 hereafter) is a LMXB discovered during an outburst in 2011 \citep{Krimm2011a, Rau2011, ArmasPadilla2013c}. It is thought to contain a relatively massive BH (12.4$\pm$3.6~\Msun, \citealt{Casares2016}) in a  2.6$\pm$0.9~h orbit, one of the shortest periods reported for a BH LMXB (\citealt{MataSanchez2015a, CorralSantana2016}, see also \citealt{ArmasPadilla2023}). Its distance is estimated to be 2.3--6.3~kpc (\citealt{MataSanchez2015a,Shahbaz2013}, but see also \citealt{Charles2019b}), which implies a peak outburst luminosity of $L_\mathrm{x}$$\sim$10$^{35-36}$~\ergs \ \citep{ArmasPadilla2013c}.  
This would make J1357 a candidate for very faint X-ray binary transient, a group of LMXBs reaching  peak luminosities of \textless~10$^{36}$ \ergs \ \citep{Wijnands2006}. 

During its discovery outburst, profound dips ($\sim$0.8 mag, 50\% flux) were observed in optical wavelengths, with increasing recurrence times of $\sim$2 to $\sim$8 min as the outburst evolved (\citealt{CorralSantana2013}, hereafter \citetalias{CorralSantana2013}). These dips were suggested to be caused by vertically extended irregularities within the (outer) accretion disc, producing obscurations of the inner parts if the system is seen at high inclination (\textit{i} = $87.4^{+2.6}_{-5.6}$ deg, \citealt{Casares2022}). If that were the case, an outwards migration of the irregularities as the outburst declined may explain the increasing dip recurrence period (DRP) with time, assuming they are locked at the keplerian frequency of a particular disc annulus \citepalias{CorralSantana2013}. 

\begin{figure*}[ht!]
\centering
\includegraphics[width=\textwidth]{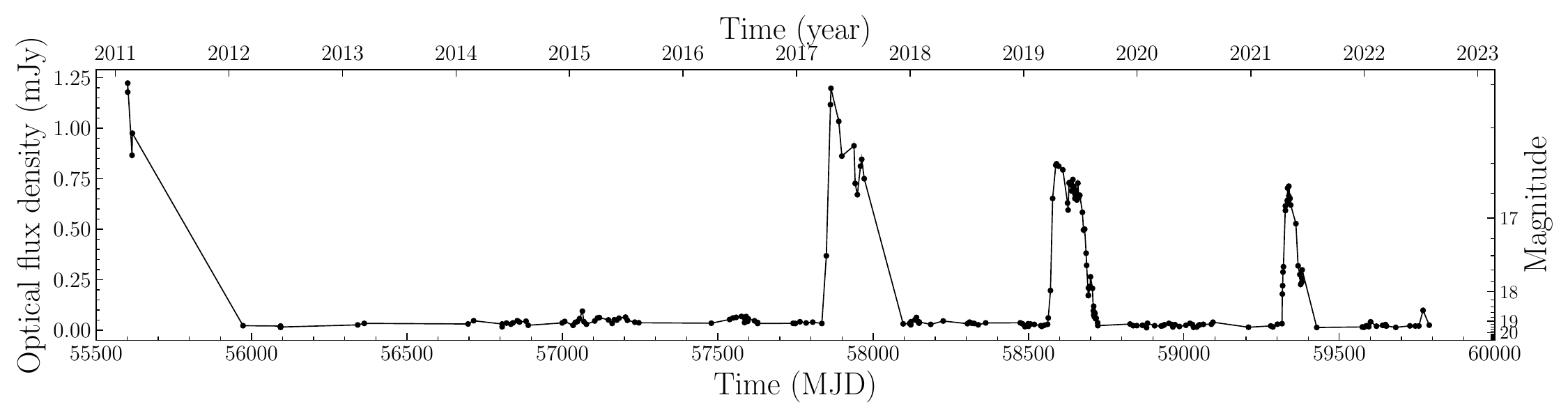}
\caption{Long-term $\textit{I}$-band flux density light curve of J1357 (Caruso et al. 2023, in prep), obtained with Las Cumbres Observatory. The 2011, 2017, 2019 and 2021 outbursts are clearly distinguished. }
\label{fig0_lc_4outb}
\end{figure*}

Analogous $\sim$0.5 mag recurrent dips were reported during the similarly bright 2017 outburst (\citealt{Drake2017, Paice2019}, \citealt{JimenezIbarra2019b}, hereafter \citetalias{JimenezIbarra2019b}). The DRP showed a temporal evolution remarkably similar to that of the discovery outburst, increasing from $\sim$2 to $\sim$5 min as the event progressed. In this occasion, a possible connection between the dips and the presence of outflows was found based on the observation of deep and broad blue-shifted absorption components in the Balmer and helium lines, concurrent with the dips
(\citealt{Charles2019b}, \citetalias{JimenezIbarra2019b}). Despite the high inclination of the system, no X-ray dips have been observed. Quasi-periodic oscillations at frequencies similar to those of the optical DRPs have been reported \citep{ArmasPadilla2014, Beri2023}. However, it has not been conclusively determined if they are produced by the same physical phenomena. 

J1357 also showed renewed outburst activity in 2019 and 2021 \citep{vanVelzen2019, Bellm2021}, although at  significantly lower luminosities ($\textit{I}$$\sim$16.6-16.8, compared to peak magnitudes of $\textit{I}$$\sim$16.2-16.1 in the first two events, \citealt{Caruso2021}). To date, there are no studies of the optical dips during the 2021 outburst, while preliminary results were only briefly reported for the 2019 event dips ($\sim$0.4~mag and DRP$\sim$11~min, \citealt{JimenezIbarra2019c}). 
 
In this paper we present time-resolved photometry of J1357 during its two most recent outbursts, 2019 and 2021, with particular emphasis on the search and characterization of optical dips. We also perform a comprehensive analysis of the dips and their observational properties over the four outbursts. We complement this work with a global X-ray study, including the first analysis of the 2021 dataset. 

\section{Observations and data reduction}\label{cap.observations}

\subsection{Optical photometry}\label{cap.observations.optical}
We performed high-time resolution photometric observations of J1357 using the Rapid Imager for Surveys of Exoplanets (RISE, \citealt{Steele2008}) at the Liverpool Telescope (LT, \citealt{Steele2004}) in the Observatorio del Roque de los Muchachos. Table~\ref{Table_data} presents our two datasets: five epochs in 2019 (between May 26 and June 10), using a 720~nm longpass filter ($\sim$\textit{I} + \textit{Z}); and seven in 2021 (between April 19 and June 27), using the OG515 and KG3 filters ($\sim$\textit{V} + \textit{R}). Several hundred exposures of 5--15 s were obtained during each epoch  (see Table~\ref{Table_data}). The basic data reduction (bias subtraction, trimming overscan regions, and flat fielding) was performed using the automatic LT reduction pipeline\footnote{\url{https://telescope.livjm.ac.uk/TelInst/Pipelines/}}. 

In order to perform the global analysis, we also use the DRPs and magnitude results for the 2017 outburst, presented in \citetalias{JimenezIbarra2019b}. As this dataset was obtained with the same telescope and instrument than ours (OG515 and KG3 filters), we re-analysed the reduced dataset to obtain the DRP uncertainties, which are not reported in the original paper. We also use the published DRP and magnitude results of the 2011 outburst obtained by \citetalias{CorralSantana2013} with other facilities.

\begin{table}[t!]
    \addtolength{\tabcolsep}{-4pt}
    \centering
    \caption{2019 and 2021 LT observing log.}
    \begin{threeparttable}	  
    \begin{tabular}{c c c c}
        \hline
        Epoch & Start MJD (2019 Date) &  \textit{i}-magnitude & Exposures
        \\
        \hline
        \hline
        1 & 58629.87 (26 May 20:46:37)  & 17.21 $\pm$ 0.08 & 572 \texttimes \ 5 s \\ 
        2 & 58633.86 (30 May 20:35:56)	& 17.23 $\pm$ 0.08 & 347 \texttimes \ 10 s \\
        3 & 58639.86 (05 Jun 20:40:07) 	& 17.19 $\pm$ 0.06 & 400 \texttimes \ 10 s \\
        4 & 58641.87 (07 Jun 20:56:51 ) & 17.23 $\pm$ 0.06 & 400 \texttimes \ 10 s \\
        5 & 58643.88 (09 Jun 21:04:46) 	& 17.23 $\pm$ 0.06 & 400 \texttimes \ 10 s \\
        \hline
        \\
    \end{tabular}
    \begin{tabular}{c c c c}
        \hline
        Epoch & Start MJD (2021 Date) &  \textit{r}-magnitude & Exposures
        \\
        \hline
        \hline
        1 & 59323.97 (19 Apr 23:22:32)  & 17.31 $\pm$ 0.09 & 720 \texttimes \ 5 s \\ 
        2 & 59325.95 (21 Apr 22:53:50)	& 17.25 $\pm$ 0.04 & 720 \texttimes \ 5 s \\
        3 & 59336.00 (01 May 23:53:27) 	& 17.14 $\pm$ 0.05 & 1010 \texttimes \ 10 s \\
        4 & 59342.95 (08 May 22:47:25) 	& 17.20 $\pm$ 0.03 & 505 \texttimes \ 10 s \\
        5 & 59361.96 (27 May 23:07:47) 	& 17.57 $\pm$ 0.07 & 384 \texttimes \ 10 s \\
        6 & 59366.93 (01 Jun 22:23:29) 	& 17.83 $\pm$ 0.04 & 384 \texttimes \ 10 s \\
        7 & 59391.93 (26 Jun 22:19:20)	& 19.61 $\pm$ 0.30 & 260 \texttimes \ 15 s \\ 
        \hline
    \end{tabular}
\end{threeparttable}
\label{Table_data}
\end{table}

\begin{figure*}
\centering
\begin{subfigure}{\columnwidth}
    \includegraphics[width=\columnwidth]{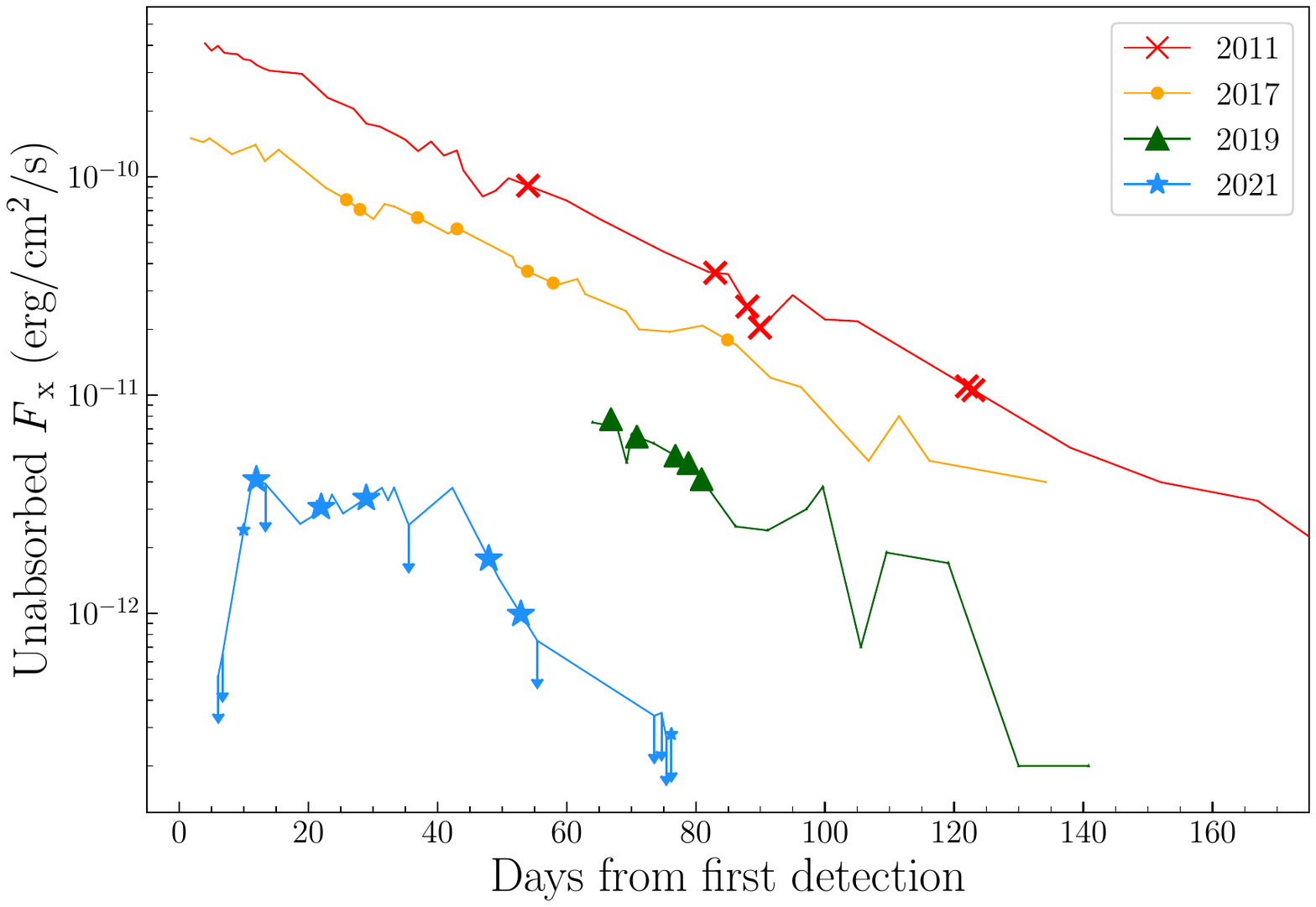}    
\end{subfigure}
\begin{subfigure}{\columnwidth}
    \includegraphics[width=\columnwidth]{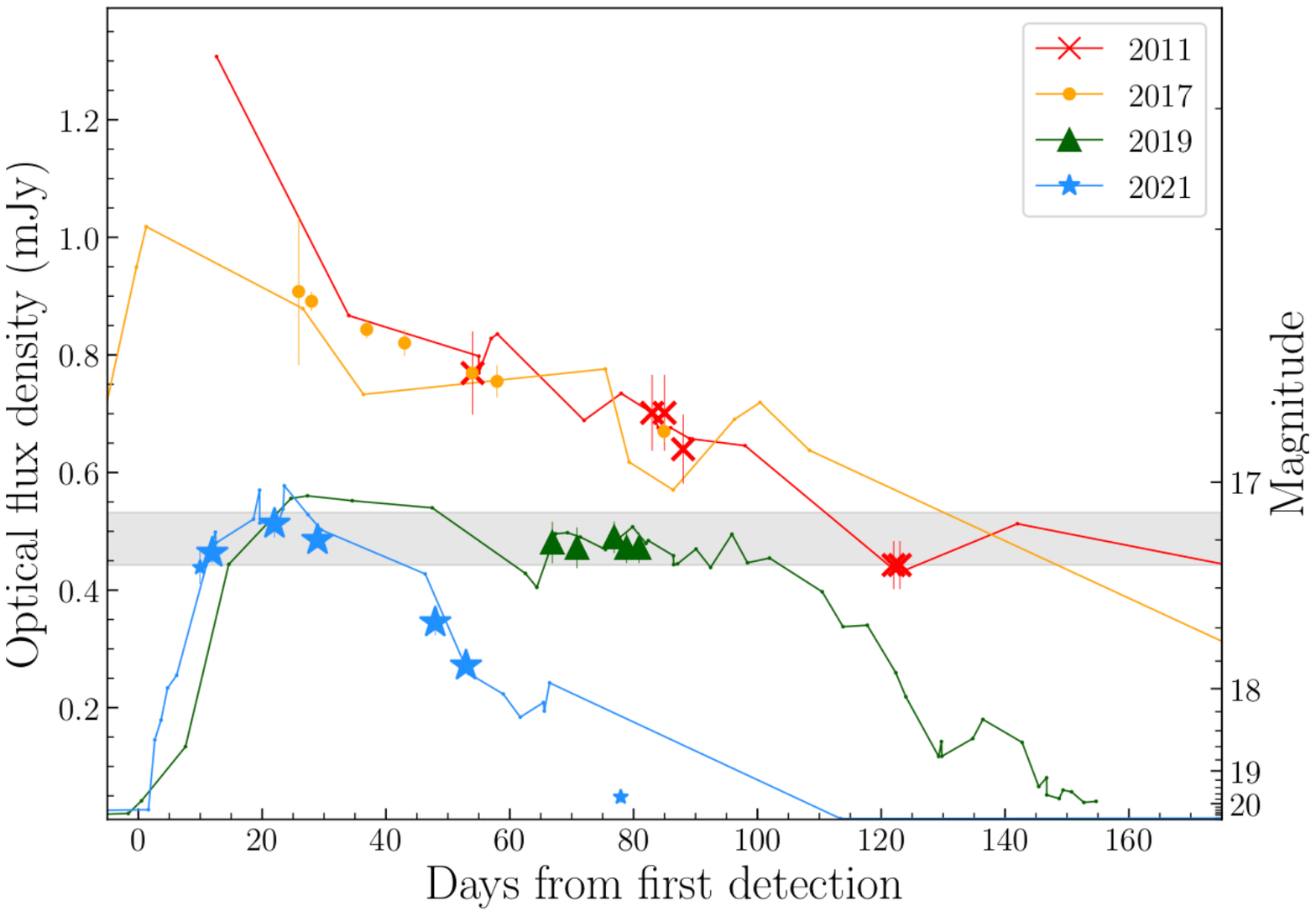}
\end{subfigure}
\caption{X-ray (\textit{Swift} and NICER, left panel) and optical (LCO, right panel) light curves of the four outbursts (solid lines). In both panels, the photometric epochs used in this work are indicated by symbols (crosses, dots, triangles and stars for 2011, 2017, 2019 and 2021 outbursts, respectively), with smaller markers for epochs with no significant DRPs (\#1 and \#7 in 2021, see Sect.~\ref{cap.analysis.peculiarities}).  The dates of first detections are given in Table~\ref{table_dips}. The grey band indicates an optical magnitude interval (17.1–17.3) with peculiar DRP evolution (see Fig.~\ref{fig5_freqVsMag}, and Sect.~\ref{sect.DRPoptFluxDep}).} 
\label{fig1_light}
\end{figure*}

\subsection{X-ray spectra}\label{cap.observations.xr}
We analysed the observations obtained during the 2021 outburst with the X-Ray Telescope (XRT, \citealt{Burrows2005}) onboard the Neil Gehrels \textit{Swift} Observatory (\textit{Swift}, \citealt{Gehrels2004}). Out of the 27 available observations, we used the 21 with a source detection. All of them were obtained in Photon Counting (PC) mode from MJDs 59319 to 59390 (0.6--1.6~ks each, see Table~\ref{Table_Rx}).

The spectra were reduced using the \textsc{heasoft} v. 6.29 package with the \textsc{xrtpipeline} (v. 0.13.6). Following the criteria of \citet{ArmasPadilla2013c}, source spectra and light curves were extracted with \textsc{xselect} (v. 2.4) using a circular region of 23.6 arcsec (10 pixels), while the background was extracted from three nearby circular regions of the same size. The response matrix files (RMFs) were obtained from the HEASARC database, while the exposure maps and ancillary response files (ARFs) were created using \textit{Swift} standard analysis threads\footnote{\url{https://swift.gsfc.nasa.gov/analysis/xrt_swguide_v1_2.pdf}}. Finally, we used \texttt{grppha} to group the spectra to a minimum of 5 photons per bin  due to the low number of counts during this outburst. We note that the $\chi^2$ minimisation technique has been found to be still valid with so few photons (e.g., \citealt{Wijnands2002, ArmasPadilla2013c, Wijnands2013}). Our results using $\chi^2$ statistics are consistent with those obtained using C-statistics. 

In addition, we make use of the \textit{Swift}/XRT 0.5--10 keV X-ray fluxes already published for the 2011, 2017 and 2019 outbursts \citep{ArmasPadilla2013c,Beri2019b,Beri2023}. We also use some published Neutron star Interior Composition Explore (NICER, \citealt{Gendreau2012}) X-ray fluxes from the 2019 outburst \citep{Beri2023}.

\section{Analysis and results}\label{cap.analysis}

Figure \ref{fig0_lc_4outb} presents the long-term \textit{I}-band light curve of J1357 from Las Cumbres Observatory (LCO, Caruso et al. 2023, in prep; see also \citealt{Russell2018,Pirbhoy2019,Goodwin2020} for details). These observations are part of an on-going monitoring campaign of $\sim$50 LMXBs coordinated by the Faulkes Telescope Project \citep{Lewis2008}. The dataset reveals that the most recent outbursts (2019 and 2021) were fainter and shorter than in 2011 and 2017. Our photometric observations and X-ray light curves, presented in Fig.~\ref{fig1_light}, confirm this difference. The optical light curves (right panel) were obtained in \textit{I}-band except for 2011 (\textit{R}-band, \citealt{Russell2018}). For visual purposes, we re-scaled them by factors of 0.85, 0.65 and 0.81 for 2017, 2019 and 2021, respectively. The X-ray light curves (left panel in Fig.~\ref{fig1_light}) were obtained with \textit{Swift} (\citealt{ArmasPadilla2013c, Beri2019b, Beri2023}; this work), except for the three first data points of the 2019 X-ray light curve, which were obtained with NICER \citep{Beri2023}.
 
\subsection{X-ray light curve analysis}\label{cap.analysis.XRlightcurves}
We fitted the X-ray spectra using \textsc{xspec} (v.12.0),  assuming a constant column density of \textit{N}$_{\mathrm{H}}$ = 1.2 $\times$ 10$^{20}$ cm$^{−2}$ \citep{ArmasPadilla2014b, Krimm2011a}. All 2021 observations were  fitted with a simple power law (\texttt{powerlaw}) affected by photoelectric absorption (\texttt{phabs}). In observations with not enough counts to perform spectral fits, we computed the 95\% confidence upper limits on the flux  following \citet{Gehrels1986} and using WebPIMMS \citep{Mukai1993}. In those cases, we fix the photon index ($\Gamma$) to that obtained in the closest observation with a valid fit, and compute the X-ray fluxes in the 0.5--10~keV band (see Table~\ref{Table_Rx}).

\subsection{Optical light curve analysis}\label{cap.analysis.lightcurves}
Our optical datasets were obtained during a brief period  of the 2019 outburst ($\sim$15 days) while covering most of the 2021 episode ($\sim$68 days from the rising phase to the end of the event, plus a later epoch following a short rebrightening,  \citealt{Baglio2021}).

We carried out aperture photometry using the \textsc{hipercam} pipeline \citep{Dhillon2018}. We performed the flux calibration against nearby stars from the PanSTARRS DR2 catalogue \citep{Chambers2016} in the most similar filters (\textit{i} and \textit{r} for 2019 and 2021, respectively), following the criteria of \citetalias{JimenezIbarra2019b}. 

\begin{figure*}[ht!]
\centering
\includegraphics[width=\textwidth]{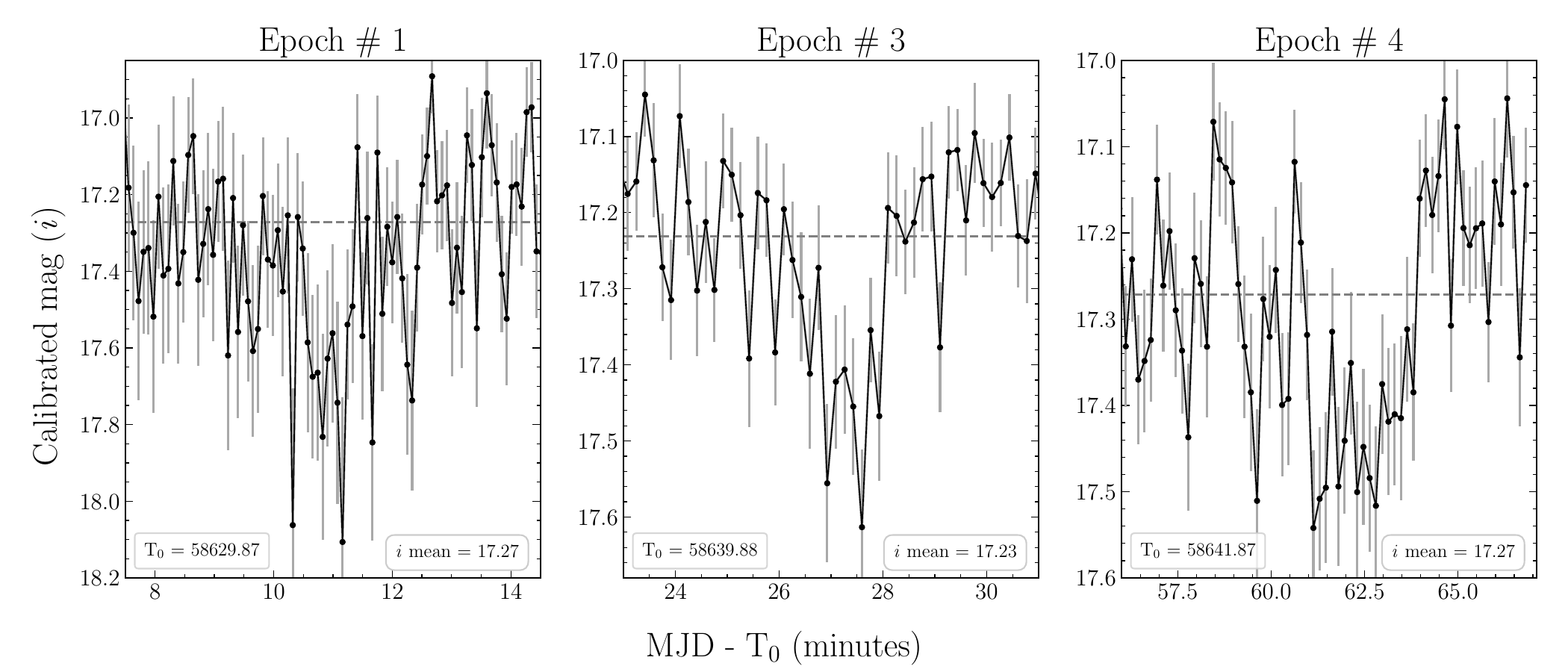}
\caption{Detail of three optical dips observed during  the 2019 outburst. The horizontal dashed line indicates the mean magnitude of each epoch. The starting MJD and mean \textit{i}-mag of each epoch are indicated in lower left and right boxes, respectively.}
\label{fig2c_dips19}
\end{figure*}

\begin{figure*}[h!]
\centering
\includegraphics[width=\textwidth]{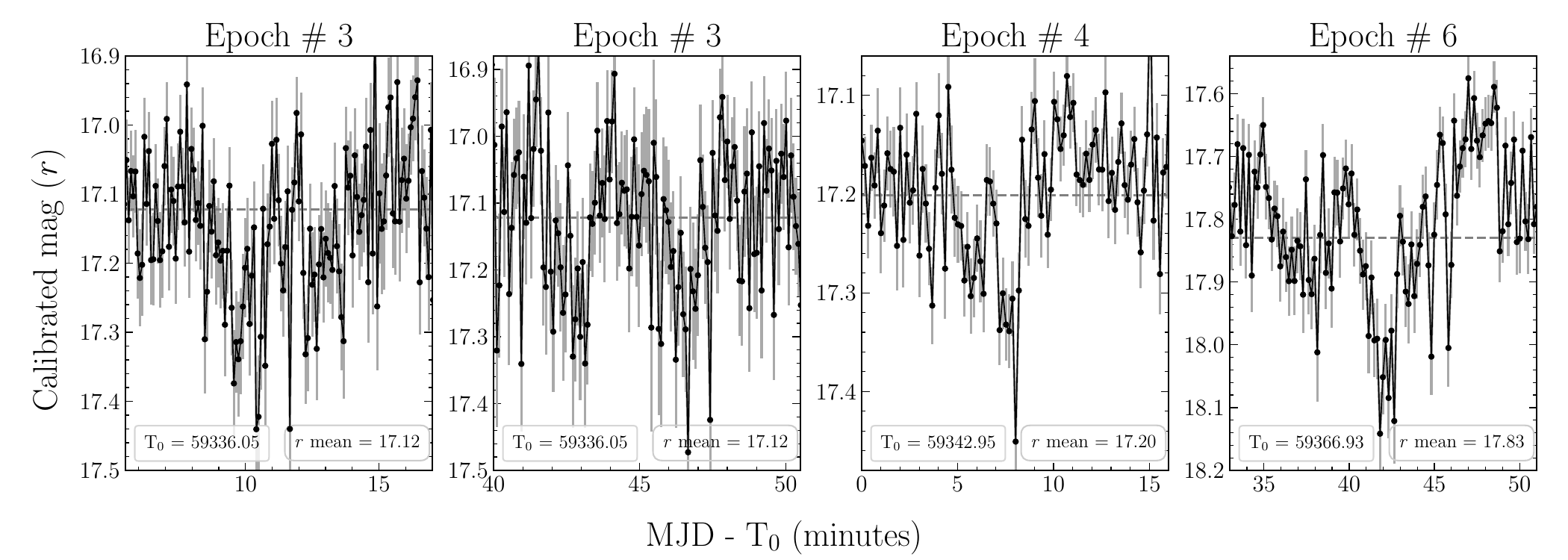}
\caption{Detail of four optical dips observed during the 2021 outburst. The horizontal dashed line indicates the mean magnitude of each epoch. The starting MJD and mean \textit{r}-mag of each epoch are indicated in lower left and right boxes, respectively.}
\label{fig2c_dips21}
\end{figure*}

\begin{figure*}
\centering
\includegraphics[width=18truecm]{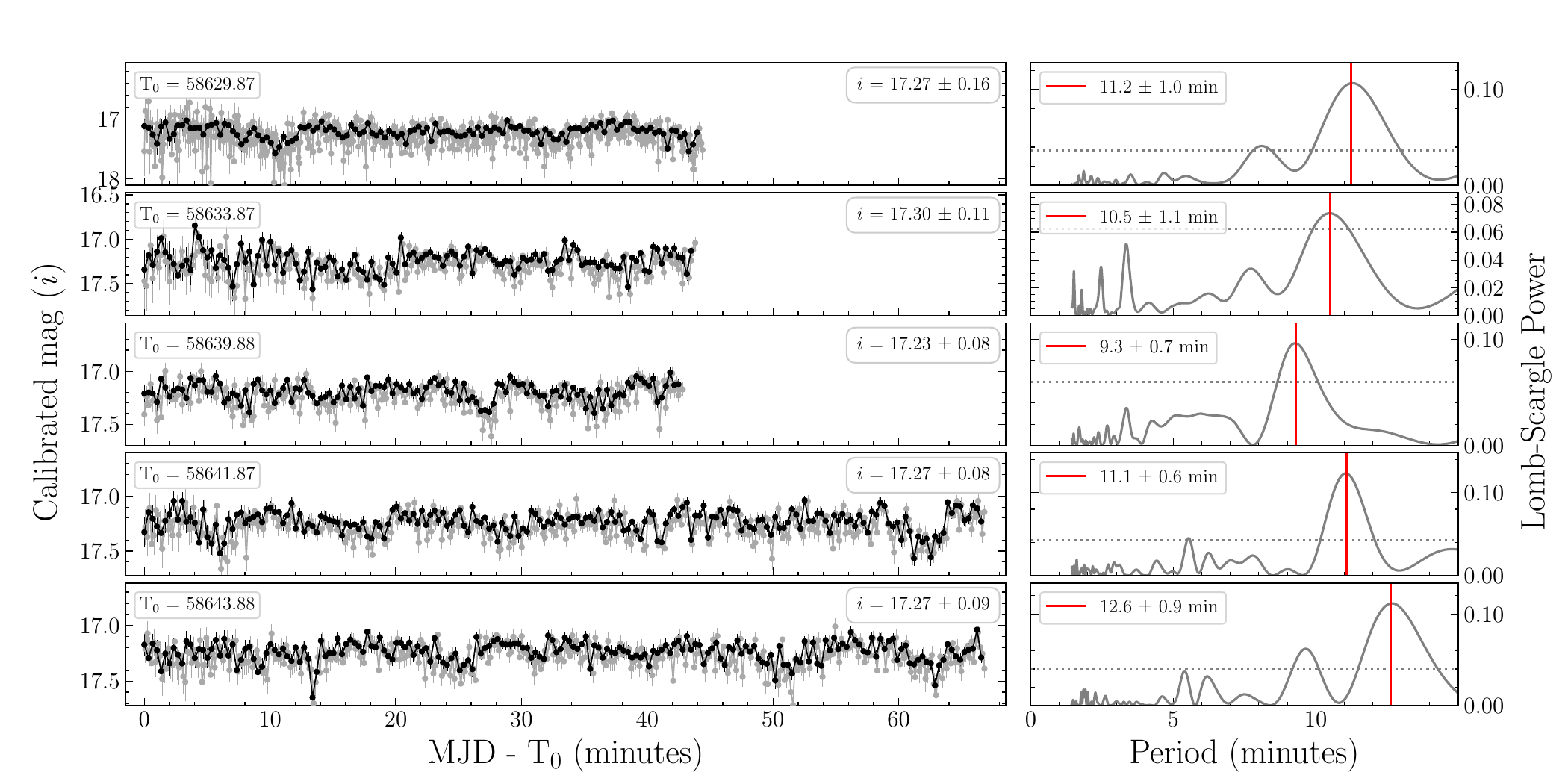}
\caption{From top (epoch \#1) to bottom (\#5), optical light curves (left) and periodograms (right) for the 2019 outburst. Light curves averaged into 20~s bins are superimposed for clarity (black). The starting MJD and mean magnitude of each epoch are shown in upper left and right boxes, respectively. The 0.2 FAP (80\% significance) and the highest, significant peak are indicated in the periodograms with a horizontal, dashed line and a vertical red line, respectively.}
\label{fig2a_periodograms2019}
\includegraphics[width=18truecm]{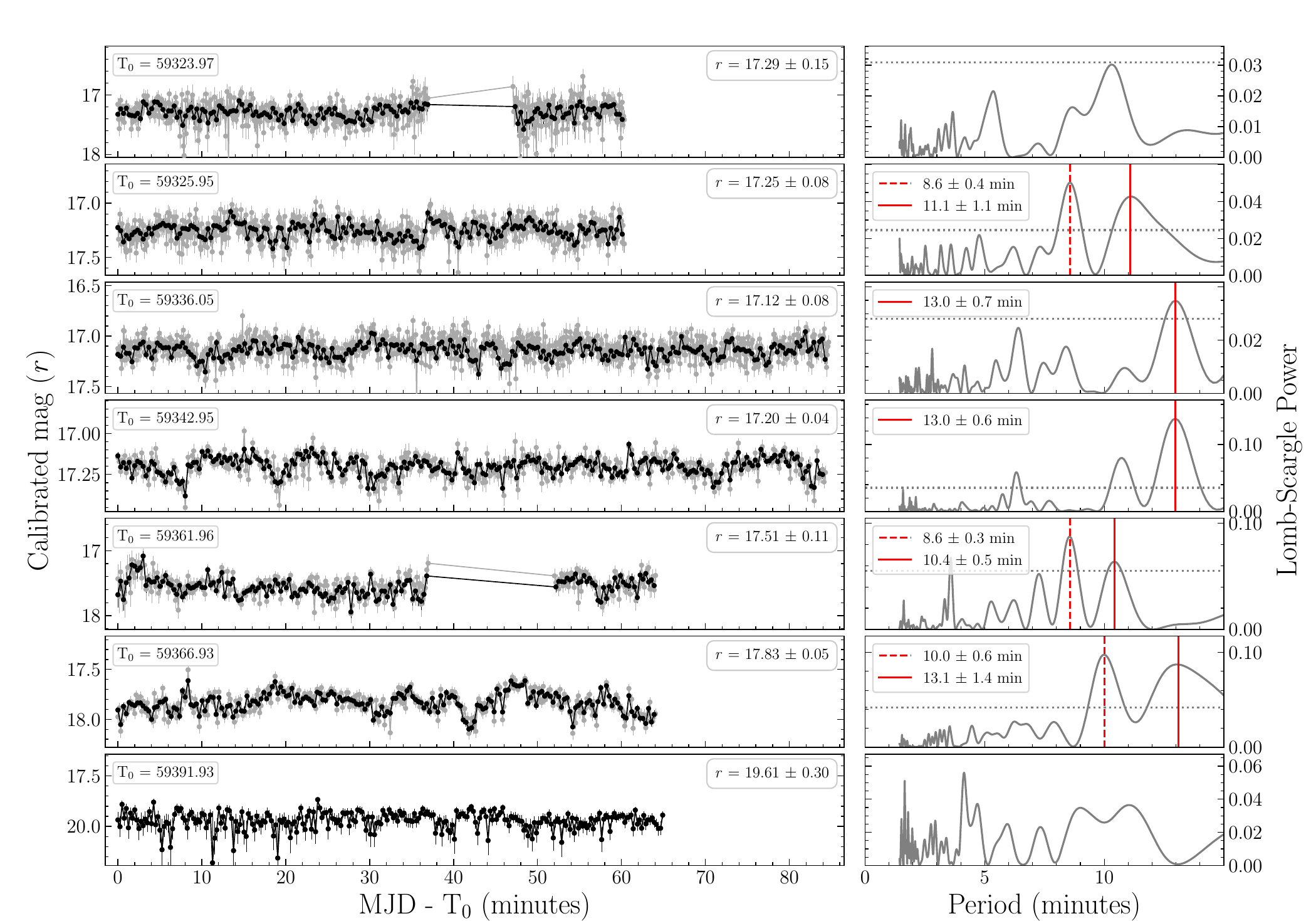}
\caption{From top (\#1) to bottom (\#7), optical light curves (left) and periodograms (right) for the 2021 outburst, using the same colour and line codes than Fig.~\ref{fig2a_periodograms2019}. We note that epochs \#1 and \#7 do not show significant peaks in the LSPs.  When the favoured DRP does not correspond to the peak with the highest power (see Sect.~\ref{cap.analysis.evolution}), the last one is indicated with a vertical, dashed line (\#2, \#5 and \#6). Epoch \#7 is not averaged into 20~s bins as the exposure time was 15~s.}
\label{fig2b_periodograms2021}
\end{figure*}

We used these high cadence observations to construct detailed light curves of each epoch and found a $\sim$0.5~mag variability level and a standard deviation of $\sim$0.05~mag during both outbursts. A visual, qualitative inspection revealed dips in both events, producing drops of $\sim$0.2--0.3~mag during $\sim$2--3~min. Some examples are shown in Figs.~\ref{fig2c_dips19} and \ref{fig2c_dips21}, while the complete light curves of each epoch are presented in the left panels of Figs.~\ref{fig2a_periodograms2019} and \ref{fig2b_periodograms2021}.


\begin{figure}
\centering
\includegraphics[width=\columnwidth]{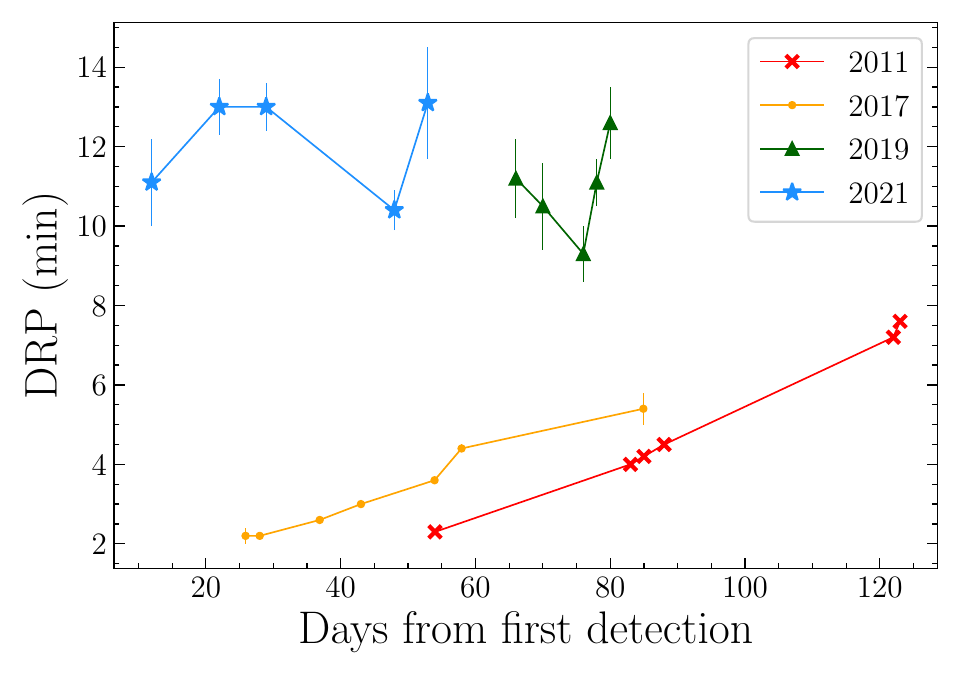}
\caption{DRP evolution with time during the four outbursts of J1357. Colours and symbols are the same than in Fig.~\ref{fig1_light}.}
\label{fig5b_freqVsTime}
\end{figure}
\begin{figure*}
\centering
\begin{subfigure}{\columnwidth}
    \includegraphics[width=\columnwidth]{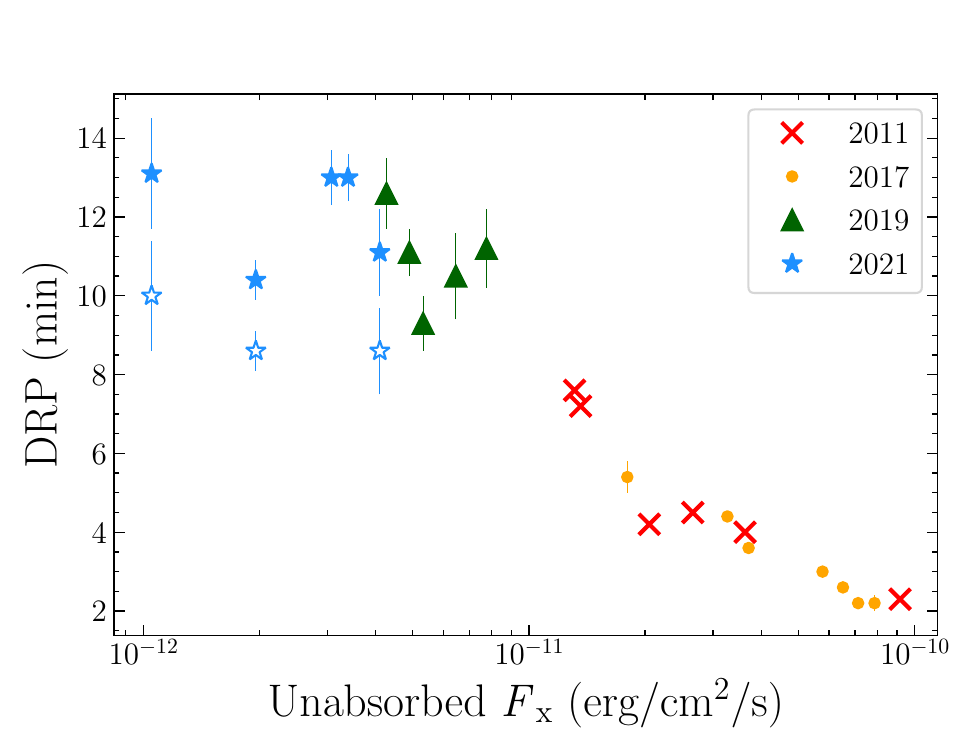}
\end{subfigure}  
\begin{subfigure}{\columnwidth}
     \includegraphics[width=\columnwidth]{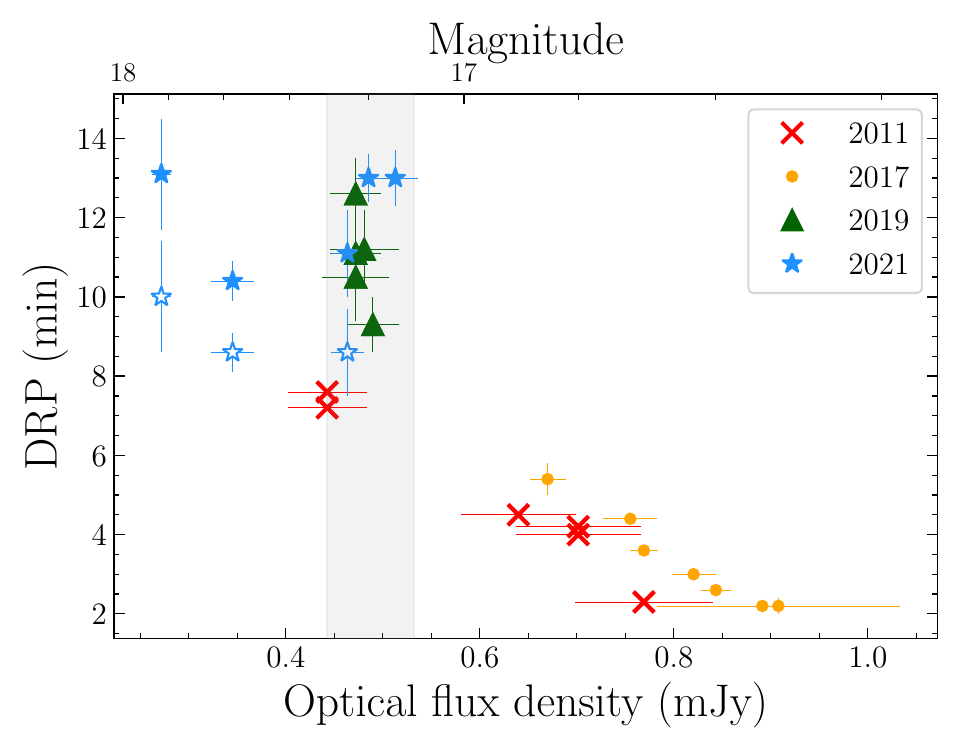} 
\end{subfigure}
\caption{DRP evolution with X-ray (left) and optical (right) fluxes for the four outbursts of J1357. The X-ray flux for each DRP was interpolated between those of the closest X-ray observations (0-3 days). The 2017, 2019 and 2021 LT datasets only include epochs with significant peaks in the LSPs (see Figs.~\ref{fig2a_periodograms2019}, \ref{fig2b_periodograms2021} and  \ref{figA1_2017}). Open markers (epochs \#2, \#5 and \#6 of 2021) indicate the DRPs corresponding to the LSP peaks with the highest power (Sect.~\ref{cap.analysis.evolution}). The grey band indicates an optical magnitude interval (17.1–17.3) with a peculiar DRP evolution (also marked in Fig.~\ref{fig1_light}, see Sect.~\ref{sect.DRPoptFluxDep}).} 
\label{fig5_freqVsMag}
\end{figure*}

\subsection{Periodogram analysis}\label{cap.analysis.peculiarities}
The right panels of Figs. \ref{fig2a_periodograms2019} and \ref{fig2b_periodograms2021} show the Lomb-Scargle Periodograms (LSP) for the 2019 and 2021 datasets, respectively. We used the \textsc{lomb–scargle python} class \citep{VanderPlas2018} as described in \citetalias{JimenezIbarra2019b}. Both datasets were detrended before the analysis using a linear fit to avoid low--frequency components. We consider only frequencies between 1.5 and 15~min in order to reduce contamination effects from the observational window  (5--15~s exposures for 70--90~min). As the main periodic features of this source were associated to optical dips in previous outbursts  \citepalias{CorralSantana2013, JimenezIbarra2019b}, we assume that periodicities in our LSPs would also be produced by similar optical dips (see Figs.~\ref{fig2c_dips19} and \ref{fig2c_dips21}), that is, the peaks of our LSPs are tracking the DRP.

We estimated the DRP uncertainty as the sigma of a Gaussian fit to the peak in the LSP \citep{VanderPlas2018}. However, this estimation is based on the assumption that the chosen peak is the correct one, which could introduce additional errors in some cases.
In order to study the validity of detected peaks, we estimated their significance by calculating the false alarm probability (FAP) at a given level of significance. 
The FAP of a peak is the probability that a dataset with no periodic signal produces a peak with similar intensity \citep{VanderPlas2018}. Thus, given a chosen FAP, we can compute the required peak height to reach such value and use it as a significance limit for our periodogram: peaks higher than the limit would be true with 1-FAP confidence.
We computed FAPs of 0.2 (80\% confidence) by the bootstrap method\footnote{We used the routine \texttt{false$\mathunderscore$alarm$\mathunderscore$level} (with method = ``bootstrap'') from  astropy.timeseries/LombScargle.}. 
Despite being arguably the most robust method for FAP estimations, it can be computationally expensive. We tried 100, 1000, and 10000 resamplings and found 1000  to be a good compromise.  Although we present the results of this analysis for every epoch (Figs. \ref{fig2a_periodograms2019} and \ref{fig2b_periodograms2021}), we will only discuss those with significant peaks, which excludes two epochs of the 2021 dataset (\#1 and \#7). Adopting a more strict FAP threshold of 90\% would exclude one additional epoch (\#3 in 2021, see Sect.~\ref{cap.analysis.evolution}). However, as visual inspection reveals clear dips in this epoch (see Fig.~\ref{fig2c_dips21}), we decided to keep the 80\% threshold.

We also re-analysed the 2017 dataset presented in \citetalias{JimenezIbarra2019b} to estimate the DRPs uncertainties (see Appendix~\ref{cap.appendix2017} for details). Our DRPs are in agreement with those reported by \citetalias{JimenezIbarra2019b} in every epoch, and all of them are significant according to our 0.2 FAP criteria. We note that a FAP threshold of 90\% would exclude epochs \#1 and \#6.

\subsubsection{DRP evolution }\label{cap.analysis.evolution}

All the epochs in the 2019 dataset show significant frequency peaks in the LSP (Fig.~\ref{fig2a_periodograms2019}): 11.2$\pm$1.0~min, 10.5$\pm$1.1~min, 9.3$\pm$0.7~min, 11.1$\pm$0.6~min and 12.6$\pm$0.9~min in epochs \#1 to \#5, respectively. 
Our analysis does not reveal a clear DRP trend with time in this dataset, which was obtained in a very short window (15 days) compared to that of previous studies (70 and 60 days during the 2011 and 2017 outbursts, respectively, see Fig.~\ref{fig1_light}).

Our  2021 dataset, on the other hand, was obtained over a much more extended timeframe, comparable to that of the first two outbursts. Moreover, our earliest observation was performed during the rising phase of the outburst (10 days after the initial detection) and hence, comparatively earlier than in 2011 and 2017 (54 and 25 days after their first X-ray detections, respectively; \citealt{Krimm2011a, Drake2017}). Five of the seven 2021 epochs show significant frequency peaks in the LSPs (all but epochs \#1 and \#7\footnote{We note that the epoch \#1 peak at $\sim$10~min is significant at 70\%. On the other hand, epoch \#7 was obtained during the final decay of the outburst, when the flux was much lower than in other epochs.}, see Fig.~\ref{fig2b_periodograms2021}). Two of them, epochs \#3 and \#4, show clear maximum peaks at 13~min. These strong LSP peaks can be confidently associated with DRPs due to the long coverage available for these epochs ($\sim$85~min). However, the other three (\#2, \#5 and \#6) show pairs of significant, close peaks of similar power: 8.6$\pm$0.4~min and 11.1$\pm$1.1~min in epoch \#2,  8.6$\pm$0.3~min and 10.4$\pm$0.5~min in \#5 and 10.0$\pm$0.6~min and 13.1$\pm$1.4~min in \#6. As their shorter coverage ($\leq$~65~min) could bias our method towards lower DRPs, we discuss both peaks, but favouring frequencies closer to those found in the epochs with longer coverages (\#3, \#4) instead of simply choosing the peaks with the highest power. We note that a similar discussion is not necessary in 2017 and 2019 datasets, as other significant peaks show considerably less power or can be interpreted as aliases of the peak with highest power (see epochs \#2--5 and \#7 in 2017, and \#1 and \#5 in 2019,  Figs.~\ref{figA1_2017} and \ref{fig2a_periodograms2019}, respectively). 

Given that some of the observed optical dips are considerably less prominent that those of the 2011 and 2017 outbursts, we performed an additional test of their reliability. We studied the light curves of a field star of similar brightness and performed the same LSP analysis. We found no similar optical dips in the light curves, nor significant recurrent periods in the periodograms according to our FAP criteria.

Overall, the 2019 and 2021 DRPs are considerably longer than those of the 2011 and 2017 outbursts \citepalias{CorralSantana2013, JimenezIbarra2019b}.  
Furthermore, they do not show the clear and unique increasing trend with time observed in the first two outbursts (see Sect.~\ref{cap.discussion_DRPvariation} and Fig.~\ref{fig5b_freqVsTime}).

\begin{table*}[ht!]
    \addtolength{\tabcolsep}{-3pt}
    \centering
    \caption{Dip properties as a function of the J1357 outburst.}
    \begin{threeparttable}	  
    \begin{tabular}{c c c c c c c c c c }
        \hline
        Year & State and & Depth\tnote{\S} &  Duration & DRP & Span of obs. & Peak \textit{I}-band & Peak \textit{F}$_{opt}$& Peak \textit{L}$_{\mathrm{x}}$\tnote{\textdagger} &  References \\
        & first detection & (mag) & (min) & (min) & (days) & (mag) & (mJy) & (erg/s) &         \\
        \hline
        \hline
        2011 & Outburst (55589) & $\sim$0.8      & $\sim$2   & 2.3--7.5  & 69 & 16.2 & $\sim$1.2 & $\sim$1 $\times$ 10$^{36}$  & 1, 2, 7 \\ 
        2017 & Outburst (57863) & $\sim$0.5      & $\sim$2   & 2.1--5.4  & 59 & 16.1 & $\sim$1.3 & $\sim$6 $\times$ 10$^{35}$  & 2, 4, 7, 8, this work \\
        2019 & Outburst (58563) & $\sim$0.3      & $\sim$2   & 9--13     & 15 & 16.6 & $\sim$0.8 & $\sim$3 $\times$ 10$^{34}$  & 2, 5, 7, this work\\
        2021 & Outburst (59314) & $\sim$0.2      & $\sim$2   & 9--13     & 68 & 16.8 & $\sim$0.7 & $\sim$1 $\times$ 10$^{34}$ &  6, 7, this work\\
        2012 & Quiescence       & $\sim$0.5-1.0  & $\sim$2   & 21.5      & 2  & 21.5\tnote{\textdaggerdbl} & $\sim$0.01  & $\sim$1.4 $\times$ 10$^{31}$ &  3, 9, this work \\ 
        \hline
    \end{tabular}
\begin{tablenotes}
\item[\S] Obtained in \textit{r}-band except for 2019 (\textit{i}-band; see Sect.~\ref{cap.observations.optical}).
\item[\textdaggerdbl] Here we refer to an approximate quiescence value in \textit{r}-band.
\item[\textdagger] Assuming a distance of 6 kpc \citep{ArmasPadilla2014b}.
\item[] References: (1) \citetalias{CorralSantana2013}, (2) \citet{Russell2019d}, (3) \citet{Shahbaz2013}, (4) \citetalias{JimenezIbarra2019b}, (5) \citet{JimenezIbarra2019a}, (6) \citet{Baglio2021}, (7) \citet{Caruso2021}, (8) \citet{Beri2019b}, (9) \citet{ArmasPadilla2014b}.
\end{tablenotes}
\end{threeparttable}
\label{table_dips}
\end{table*}

\subsubsection{DRP dependence with flux}\label{cap.analysis.fig5}

To explore a possible dependence between the DRP and X-ray and optical fluxes, we plotted both variables for every outburst in Fig.~\ref{fig5_freqVsMag}. The DRP-\textit{F}$_\mathrm{x}$ plot (left panel of Fig.~\ref{fig5_freqVsMag}) suggests two main results: (i) the DRPs are longer for fainter outbursts and (ii) in general, they seem to increase as the outburst decays and the flux drops. Therefore, a DRP-flux dependence seems probable. We perform a Pearson correlation to test this possible linear correlation and find a -0.86 coefficient, which supports its existence (p-value 2$\times$10$^{-8}$).

It is relevant to note the case of epochs \#2, \#5 and \#6 in the 2021 dataset, where the peaks with the highest LSP power (open stars in Fig.~\ref{fig5_freqVsMag}) point to lower DRPs than the overall trend might suggest. As discussed above (Sect.~\ref{cap.analysis.evolution}), for these epochs we favour the significant peaks found at longer DRPs and slightly less LSP power (coloured stars), which produce a significantly less scattered (i.e. better defined) correlation at low fluxes. In any case, the scatter of the correlation increases as we move to lower fluxes and longer DRPs. This is expected as the signal-to-noise becomes lower, the dips shallower and the DRPs approach values closer to our observing window ($\sim$15~min DRPs for 60~min windows). A Pearson correlation test using the peaks of highest power for those epochs (\#2, \#5 and \#6) results in a coefficient of -0.82 (p-value 9$\times$10$^{-7}$).

This general correlation of DRPs increasing as flux decreases 
seems to hold also when we use optical instead of X-ray fluxes (right panel of Fig.~\ref{fig5_freqVsMag}), with a Pearson coefficient of -0.89 with p-value 1$\times$10$^{-8}$ (-0.88 when using the values corresponding to open markers for epochs \#2, \#5 and \#6, with p-value 1$\times$10$^{-7}$). However, we note that the scatter is particularly noticeable within a narrow flux band (17.1--17.3~mag, grey region in Fig.~\ref{fig5_freqVsMag}). This band coincides with the optical peaks of the 2019 and 2021 outbursts, while the rest of the points of the correlation occur during the final decay (see Fig.~\ref{fig1_light}).

Finally, we stress that the correlation would hold even if we chose a  FAP threshold of 90\% (see Sect.~\ref{cap.analysis.peculiarities}).

\section{Discussion}\label{cap.discussion}
Since its discovery, J1357 has gone into outburst four times in ten years. The first two, 2011 and 2017, were first noticed in X-rays and showed very similar luminosity peaks, temporal evolution and optical dips. The following outbursts, 2019 and 2021, were discovered at earlier stages by optical facilities. Thus, earlier observations and better coverage of their optical peaks were possible. Our analysis reveals optical, recurrent dips  during these two fainter and shorter outbursts, demonstrating that they have been present in every observed outburst of the source. Thus, regardless of the luminosity and duration of the outburst, these optical, recurrent dips are most likely characteristic of this system. Table~\ref{table_dips} summarises the properties of the dips, which are shallower and show longer DRPs in fainter outbursts despite showing similar duration ($\sim$2~min).

\subsection{The DRP evolution with time}\label{cap.discussion_DRPvariation}
Previous studies pointed to the existence of a common increasing DRP trend with time in every outburst of the source (\citetalias{CorralSantana2013, JimenezIbarra2019b}, \citealt{JimenezIbarra2019c, Paice2019}). As shown in Fig.~\ref{fig5b_freqVsTime}, this hypothesis was based on the remarkably similar DRP evolution during the decay of the 2011 and 2017 outbursts (red crosses and orange dots). 
However, this trend (i.e. steadily increase of the DRP with time as the outburst evolves) was not observed in the 2019 and 2021 events (green triangles and blue stars).  This is most likely due to the fact that the 2011 and 2017 monitoring covered the decay of the system during two very similar outbursts. The coverage of the 2019 and 2021 events was different, including the final part of the rise (2021) and the outburst peak. Also, the luminosity range that these events sample is rather different. In summary, the results presented in this work show that new variables need to be considered to characterise the evolution of the DRP across the outbursts of J1357.

\begin{figure*}[ht!]
\centering
\begin{subfigure}{\columnwidth}
    \includegraphics[width=\columnwidth]{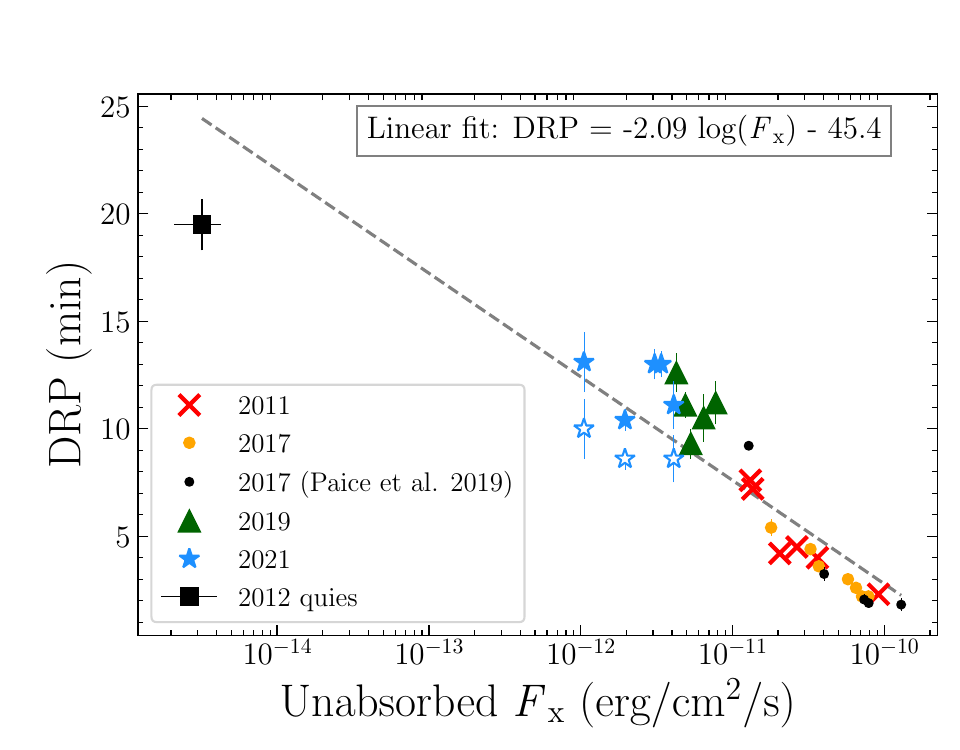}
\end{subfigure}  
\begin{subfigure}{\columnwidth}
    \includegraphics[width=\columnwidth]{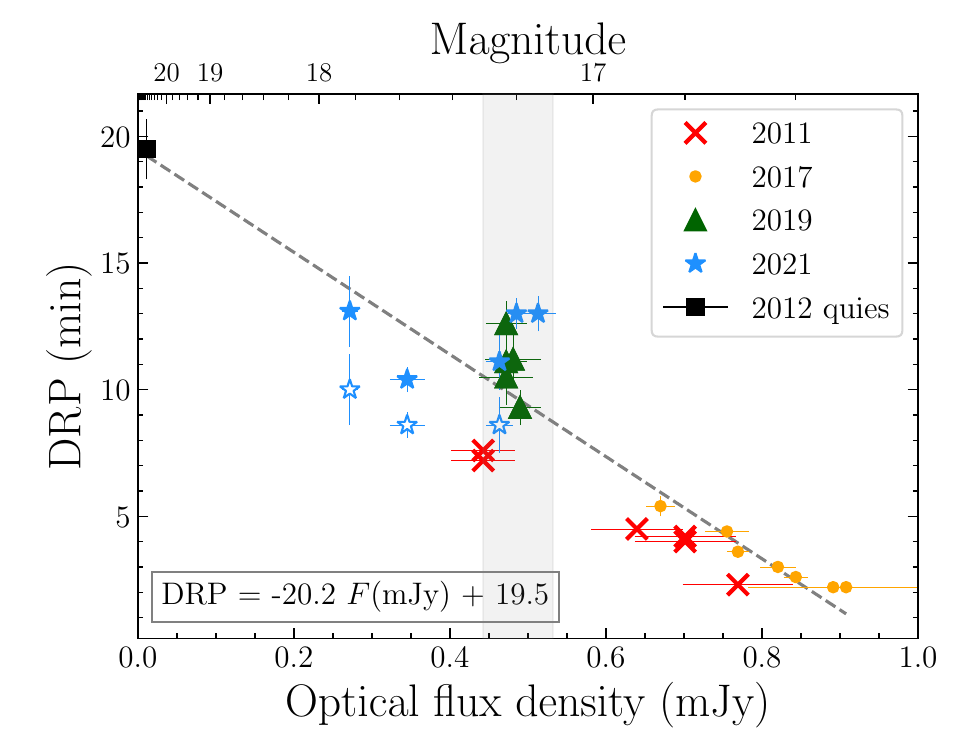}
\end{subfigure}
\caption{Same figure than Fig.~\ref{fig5_freqVsMag},  now including data from 2012 quiescence (black square) and an independent dataset from the 2017 outburst (\citealt{Paice2019}, small black dots in the left panel). Note that in the DRP-\textit{F}$_\mathrm{x}$ plot (left panel) the quiescence \textit{F}$_\mathrm{x}$ was obtained with \textit{XMM-Newton} 422 days after the DRP measurement \citep{ArmasPadilla2014b}, which could affect its position on the diagram. The coefficients of a linear fit to all the datasets (excluding 2021 open markers) are shown in a text box.}
\label{fig6_freqVsMag_plusQuies}
\end{figure*}

\subsection{A global correlation between the DRP and the observed X-ray flux}\label{sect.DRPFluxDep}
Our results reveal that bright outbursts (\textit{F}$_\mathrm{x, unabs}$$\sim$10$^{-11}$--10$^{-10}$~$\flux$) show clear dips with short DRPs ($\sim$\mbox{2--8}~min), while fainter ones (\textit{F}$_\mathrm{x, unabs}$$\sim$10$^{-12}$--10$^{-11}$~$\flux$) produce dips with shallower depths and higher DRPs ($\sim$9--13~min). Furthermore, as the X-ray flux changes through each event, the DRPs also change, increasing (in general) with fainter flux. We note that an independent optical dataset obtained during the 2017 outburst \citep{Paice2019} reported very similar DRPs that also lie in this general trend. In Fig.~\ref{fig6_freqVsMag_plusQuies} (left panel) we include all the published DRP values of the source (see also Sect.~\ref{sect.dipsQuies}).  In summary, as shown in Fig.~\ref{fig5_freqVsMag}, an overall correlation of the DRP with the X-ray flux is observed, with more prominent and frequent optical dips seen at brighter stages.

As previously explained, the scatter of this correlation increases as the X-ray flux decreases and the DRP rises (see Sect.~\ref{cap.analysis.fig5}). 
Thus, while the optical dips were easily detected in the first two outbursts, the fainter events on 2019 and 2021 would have particularly benefited from longer coverage to reduce the uncertainties in the DRP measurement. This might be taken into account in future observations.

\subsubsection{DRP evolution with the optical  flux}\label{sect.DRPoptFluxDep}
A similar DRP trend is observed with the optical flux (right panel of Fig.~\ref{fig5_freqVsMag}), reinforcing the hypothesis of a correlation between both variables. However, in this case the scatter seems to be even higher in a narrow band of optical flux during the 2019 and 2021 outbursts. This might be related to the fact that those data points were obtained during a plateau of the optical light curve (grey bands in Figs.~\ref{fig1_light}, \ref{fig5_freqVsMag} and \ref{fig6_freqVsMag_plusQuies}), while DRPs show less scatter again during the decaying phase of the 2021 outburst. It is worth noticing that although this band also contains the two last 2011 epochs, they correspond to the outburst decay and do not seem to deviate from the regular trend.

Such region of increased scatter (i.e. grey band in Figs.~\ref{fig1_light} and \ref{fig5_freqVsMag}) is not evident in the DRP-\textit{F}$_\mathrm{x}$ plot (left panel of Fig.~\ref{fig5_freqVsMag}). Assuming that the DRP-\textit{F}$_\mathrm{x}$ is the primary correlation, this additional scatter might be due to the change of the slope of the optical--X-ray flux correlation at different outburst stages observed in other systems (e.g., \citealt{LopezNavas2020}). This was not observed in the optical databases of the 2011 and 2017 outbursts, obtained during outburst decays. However,  we note that \citet{Paice2019} reported a 2017 epoch with higher DRP than expected from the general DRP-time trend found for the 2011 and 2017 datasets \citepalias{CorralSantana2013, JimenezIbarra2019b}. This epoch was observed only one week after detecting the outburst (i.e. 20 days before our first epoch and possibly during the outburst peak, see Fig.~\ref{fig6_freqVsMag_plusQuies}). In light of our results, this epoch might be affected by the same issue than those of the 2019 and 2021 events during the optical plateau at the outburst peaks.

\subsubsection{Dips in quiescence}\label{sect.dipsQuies}
Optical dips with a possible recurrence time around $\sim$30~min were reported in 2012, when the system was back in quiescence eight months after the discovery outburst \citep{Shahbaz2013}. The presence of optical dips during these dim stages (\textit{r'}$\sim$22) allows us to explore a new, different region in the DRP-flux plot. We analysed the available data (see details in Appendix~\ref{cap.appendixQuiescence}) and obtained a DRP of 21.5$\pm$0.5~min in \textit{r'} and \textit{g'} bands, which is considerably longer than those found in outburst (see Table~\ref{table_dips}). 
In Fig.~\ref{fig6_freqVsMag_plusQuies} we show how this longer recurrence time might be consistent with the main, overall correlation seen in outburst, suggesting a common physical origin for both the quiescent and outburst dips (Pearson coefficients of -0.82 and -0.92 for the DRP correlation with X-ray and optical fluxes, respectively). In this plot, we associate the quiescent DRP with the quiescent \textit{XMM-Newton} X-ray flux reported in \citet{ArmasPadilla2014b}. We will further discuss this in Sect.~\ref{sect.dipsOutflows}.

A linear fit to all the available datasets (including quiescence) was performed for the DRP correlation with X-ray and optical fluxes (see Fig.~\ref{fig6_freqVsMag_plusQuies}). This linear fit might be useful for future observations of J1357, providing a general idea of the expected range of DRPs at a given flux.

\subsubsection{Further considerations on DRP-flux dependence}
Our results must be seen with caution, as our method and datasets have some caveats.
First, the photometric filters of the LT do not have a direct correspondence with the PanSTARRS filters. Thus, our calibration might be affected by a systematic bias, which would be different in the 2019 dataset (720 nm longpass filter) with respect to 2017 and 2021 (OG515 and KG3 filters). We have estimated this bias by comparing the magnitudes obtained for 10 field stars with those reported by PanSTARRS. We find differences $\textless$~0.1~mags, with a mean offset of $\sim$~0.002 in the \textit{r}-like filter (2017 and 2021 datasets) and $\sim$ -0.05 in \textit{i}-like filter (2019 dataset). This would have no effect in the timing, but might artificially change the scatter of the DRP-flux diagrams. In any case, as this would not affect the DRP-\textit{F}$_\mathrm{x}$ (left panels of Figs.~\ref{fig5_freqVsMag} and \ref{fig6_freqVsMag_plusQuies}), which show very similar results to DRP-flux plots (right panels of Figs.~\ref{fig5_freqVsMag} and \ref{fig6_freqVsMag_plusQuies}), such systematic bias does not seem to significantly affect our conclusions.

Second, although most observations were performed using \textit{r}/\textit{r'}/\textit{R} filters, the 2019 dataset was obtained with a filter closer to \textit{i}/\textit{z}. Previous works seem to suggest that \textit{i}-mags might be $\sim$0.1-0.2~mag brighter than \textit{r}-mags for this source \citep{Baglio2021}. Thus, when comparing  outbursts in Figs.~\ref{fig1_light}, \ref{fig5_freqVsMag} and \ref{fig6_freqVsMag_plusQuies} we must be aware that the 2019 epochs might be slightly displaced with respect to the \textit{r}-band. This might alter the width of the region of increased scatter (grey band in Figs.~\ref{fig1_light} and \ref{fig5_freqVsMag}). However, even if our data suffer from such offset, it would be small enough to keep our conclusions unaltered. 
It is also relevant to note the dip-wavelength dependence found by previous studies, with deeper dips at longer wavelengths (\citealt{Paice2019}, see also \citetalias{JimenezIbarra2019b}). This might be particularly relevant for dip detection at fainter outbursts (such as those of 2019 and 2021), where we find them to be less prominent (see Table~\ref{table_dips}). Shallower dips might thus be even harder to detect in observations performed at shorter wavelengths (such as the 2021 \textit{r}-like dataset) in comparison to longer wavelengths (e.g., the 2019 \textit{i}-like dataset). However, we note that the dips are in fact detected even in our 2021 dataset, so this effect might not be particularly important in this source. 

Other possible caveats are related to the periodogram analysis. Although we were cautious in estimating uncertainties and choosing only peaks over a certain FAP, an incorrect peak choice in the LSPs would easily introduce larger errors and more scatter in the DRP-flux diagrams. This would particularly affect those periodograms with more than one peak over the FAP rate, which is the reason why we decided to include them in Figs.~\ref{fig5_freqVsMag} and \ref{fig6_freqVsMag_plusQuies}. However, we note that as previously discussed, they do not change our conclusions.

Finally, we note that there is only one quiescent DRP measurement available. Although relevant for our analysis and being consistent with our results, it might not be representative of the system behaviour during the dimmest stages. 

\subsection{Origin of the optical dips and their connection to outflows}\label{sect.dipsOutflows}
The first studies of J1357's optical dips suggested that they might be produced by irregularities in the accretion disc. These would be formed in a region that would move outwards as the outbursts evolve to fainter stages \citepalias{CorralSantana2013,JimenezIbarra2019b}. If we assume the same origin for the 2019 and 2021 optical dips, such irregularities might originate at outer locations for fainter events, as we found higher Keplerian frequencies (0.3--0.4~\Rsun, assuming a canonical BH of $\sim$10~\Msun) than those of 2011/2017 events (0.12--0.27~\Rsun \ and 0.16--0.18~\Rsun). 

Previous studies during the 2017 outburst indicated a possible connection between dips and optical outflows. These claims were based on the discovery of blue-shifted absorptions in high-time resolution optical spectra simultaneous with optical dips. These blue-shifted absorptions were not found in spectra taken outside of the dips (\citetalias{JimenezIbarra2019b}, \citealt{Charles2019b}). We note that optical winds are commonly seen in high inclination BH transients (e.g., \citealt{Munoz-Darias2016, Panizo-Espinar2022}) and in some cases a multi-phase, clumpy structure has been proposed (\citealt{Munoz-Darias2022}; see also \citealt{Motta2017}).
Thus, we can attempt to qualitatively explain our results in the context of optical dips being a direct consequence of outflows seen at high orbital inclination. As a result of the strong irradiation during outbursts, the upper layers of the disc might develop irregularities, which could produce prominent dips (as those observed in 2011 and 2017 in J1357) and clumpy equatorial winds when expelled. On the other hand, fainter outbursts with consequently lower irradiation might yield smaller irregularities at larger radii, resulting in shallower optical dips (as those observed in 2019 and 2021). 

If the dips observed in quiescence have the same physical origin as those in outburst, the possible connection between dips and outflows becomes even more puzzling. The irregularities would be located at more external radii  ($\sim$0.6~\Rsun) than those observed in outburst, but still far from the edge of the disc, which was estimated to be $\sim$1.7~\Rsun \ in outburst \citepalias{CorralSantana2013}. However, we note that accretion discs have been observed to be smaller during quiescence ($\sim$0.5~R$_{L1}$, \citealt{Casares1995, Marsh1994}, which for J1357 results in $\sim$0.95~\Rsun, \citetalias{CorralSantana2013}). During quiescence, the accretion disc is weakly irradiated, which should in principle prevent the formation of outflows via radiation-related mechanisms (e.g., \citealt{Higginbottom2020, Charles2019}).  Nonetheless, other wind launching mechanisms might be invoked, such as magnetic driven winds (e.g., \citealt{Waters2018, Tetarenko2018}). In any case, it is worth noting that to date there is no evidence of winds being launched in quiescence.
Finally, one can always speculate with the possibility that more than one mechanism contribute to create the optical dips, with different contributions dominating at different luminosities and irradiation levels.

\section{Conclusions}\label{cap.conclusions}
We have presented new photometric data of the X-ray binary transient Swift~J1357.2−-0933 during its most recent outbursts (2019 and 2021), including also the analysis of 2021  \textit{Swift}/XRT data.  Our results reveal that the optical, recurrent dips are  a common phenomena in the transient, with similar duration but different properties over different outbursts. In particular, we discover a general correlation between the dip recurrence period and the X-ray and optical fluxes, with shorter periods seen at brighter states, which might hold to some extent even in quiescence. Further optical observations of this transient during outburst to test these conclusions are encouraged. Additional quiescence data are also needed to confirm the presence of less recurrent dips at the faintest stages of the source.

Finally, it is relevant to note that this transient is singular in many aspects, with one of the highest orbital inclinations for a LMXB and a donor star whose small size would be comparable to that of the outer disc rim \citepalias{CorralSantana2013}. This might be relevant for many observational features including optical dips visibility. As a matter of fact, they might be a common feature of LMXBs, not observable at other combinations of inclination, disc thickness and companion star relative size. High-time resolution optical photometry during outbursts of other transients is  strongly encouraged to test these possibilities. 

\begin{acknowledgements}
    We are thankful to the anonymous referee for constructive comments that have improved this paper. 
    This work is supported by the Spanish Ministry of Science via an \textit{Europa Excelencia} grant (EUR2021-122010) and the \textit{Plan de Generacion de conocimiento}: PID2020-120323GB-I00, PID2020-114822GB-I00 and PID2021-124879NB-I00. DMS acknowledges support from the Spanish Ministry of Science and Innovation via an Europa Excelencia grant (EUR2021-122010). TS acknowledges financial support from the Spanish Ministry of Science, Innovation and Universities under grant PID2020-114822GB-I00. FMV acknowledges support from the grant FJC2020-043334-I financed by MCIN/AEI/10.13039/501100011033 and Next Generation EU/PRTR. This article uses material based upon work supported by Tamkeen under the NYU Abu Dhabi Research Institute grant CASS.
    \textsc{Molly} software developed by Tom Marsh is gratefully acknowledged. We acknowledge the use of public data from the \textit{Swift} data archive. The Liverpool Telescope is operated on the island of La Palma by Liverpool John Moores University in the Spanish Observatorio del Roque de los Muchachos of the Instituto de Astrof\'isica de Canarias with financial support from the UK Science and Technology Facilities Council. This work uses data from the Faulkes Telescope Project, which is an education partner of LCO. The Faulkes Telescopes are maintained and operated by LCO.
\end{acknowledgements}	

\bibliographystyle{aa}
\bibliography{J1357.bbl}

\begin{appendix}

\section{Analysis of the 2021 X-ray dataset}\label{cap.appendixRx}
We fitted the 2021 \textit{Swift}/XRT spectra of J1357 using a simple power law (\texttt{powerlaw}) affected by photoelectric absorption (\texttt{phabs}), assuming a constant column density of 1.2~$\times$~10$^{20}$ cm$^{−2}$ \citep{ArmasPadilla2014b, Torres2015}.  The results of this analysis are presented in Table~\ref{Table_Rx}.

\begin{table*}[t!]
    \addtolength{\tabcolsep}{2pt}
    \centering
    \begin{threeparttable}	  
    \caption{Log and spectral results of \textit{Swift}/XRT observations during the 2021 outburst of J1357.}
    \label{Table_Rx}
    \begin{tabular}{c c c c c c c c }
        \hline
        Obs. ID\tnote{\S} & Start date   & Exp.  & Net count rate  & $\Gamma$ & \textit{F}$_{\mathrm{X, abs}}$ \ (0.5-10 keV)\tnote{\textdagger} & \textit{F}$_{\mathrm{X, unabs}}$ (0.5-10 keV)\tnote{\textdagger} & \qhired / d.o.f.\tnote{\textdaggerdbl} \\
        &  (MJD)  &  (ks) & (counts s$^{-1}$) &  & (10$^{−12}$ $\flux$) & (10$^{−12}$ $\flux$) &  \\
        \hline
        \hline
        00031918101 & 59320.06   & 0.81 & 0.004 & 1.3 (fix) & \textless0.52 & \textless0.52 & -   \\
        00031918102 & 59320.71   & 1.63 & 0.007 & 1.3 (fix) & \textless0.65 & \textless0.65 & -   \\
        00031918104\tnote{[\#1-2]} & 59325.36   & 0.95 & 0.065 & 1.3$\pm$0.4 & 4.15$\pm$0.19 & 4.18$\pm$0.19 & 0.62/7   \\
        00031918105 & 59327.36   & 0.58 & 0.053 & 1.3 (fix) & \textless3.90 &  \textless3.92 & -    \\
        00031918106 & 59332.75   & 0.88 & 0.067 & 2.1$\pm$0.4 & 2.54$\pm$0.09 & 2.57$\pm$0.09 & 0.05/6  \\
        00031918107 & 59334.60   & 0.95 & 0.075 & 1.9$\pm$0.3 & 2.75$\pm$0.09 & 2.78$\pm$0.1 & 0.64/9   \\
        00031918108 & 59335.32   & 0.89 & 0.100 & 2.0$\pm$0.3 & 3.23$\pm$0.08 & 3.28$\pm$0.08 & 1.19/12    \\
        00031918109\tnote{[\#3]} & 59336.52   & 0.89 & 0.065 & 2.0$\pm$0.3 & 2.87$\pm$0.09 & 2.91$\pm$0.09 & 0.8/7   \\
        00031918110 & 59337.65   & 0.59 & 0.079 & 1.6$\pm$0.3 & 3.48$\pm$0.1 & 3.50$\pm$0.1 & 0.82/6   \\
        00031918112 & 59339.38   & 0.89 & 0.077 & 2.0$\pm$0.3 & 2.83$\pm$0.08 & 2.87$\pm$0.08 & 0.62/9    \\
        00031918115\tnote{[\#4]} & 59345.41   & 0.90 & 0.073 & 1.5$\pm$0.3 & 3.74$\pm$0.11 & 3.76$\pm$0.11 & 1.52/9   \\
        00031918116 & 59346.34   & 0.92 & 0.079 & 1.8$\pm$0.3 & 3.26$\pm$0.1 & 3.30$\pm$0.1 & 0.68/10   \\
        00031918117 & 59347.27   & 0.57 & 0.101 & 1.9$\pm$0.3 & 3.72$\pm$0.09 & 3.77$\pm$0.09 & 0.89/8   \\
        00031918118 & 59349.54   & 0.62 & 0.050 & 1.9 (fix) & \textless2.51 & \textless2.54 & -   \\
        00031918119 & 59356.30   & 1.01 & 0.062 & 1.8$\pm$0.3 & 3.72$\pm$0.10 & 3.76$\pm$0.10 & 0.78/8   \\
        00031918120\tnote{[\#5]} & 59363.48   & 1.07 & 0.050 & 2.2$\pm$0.5 & 1.43$\pm$0.14 & 1.45$\pm$0.13 & 1.36/5   \\
        00031918121\tnote{[\#6]} & 59369.45   & 0.85 & 0.014 & 2.2 (fix) & \textless0.74 & \textless0.75 & -    \\
        00031918122 & 59387.56   & 1.01 & 0.005 & 2.2 (fix) & \textless0.34 & \textless0.34 & -     \\
        00031918123 & 59388.69   & 1.00 & 0.005 & 2.2 (fix) & \textless0.34 & \textless0.35 & -     \\
        00031918124 & 59389.43   & 0.77 & 0.003 & 2.2 (fix) & \textless0.26 & \textless0.27 & -     \\
        00031918125\tnote{[\#7]} & 59390.22   & 0.92 & 0.003 & 2.2 (fix) & \textless0.27 & \textless0.28 & -    \\
        \hline
    \end{tabular}
\begin{tablenotes}
\item[\S] LT epochs are indicated in brackets for those X-ray epochs closest in time to the optical dataset. 
\item[\textdagger] \textit{F}$_\mathrm{X, abs}$ and \textit{F}$_\mathrm{X, unabs}$ represent the absorbed and unabsorbed X-ray fluxes (0.5-10~keV), respectively.
\item[\textdaggerdbl] \qhired \ and d.o.f. represent the reduced $\chi$$^{2}$ and the degrees of freedom, respectively.
\end{tablenotes}
\end{threeparttable}	  
\end{table*}

\section{Re-analysis of the 2017 LT dataset}\label{cap.appendix2017}
\begin{figure*}
\centering
\includegraphics[width=18truecm]{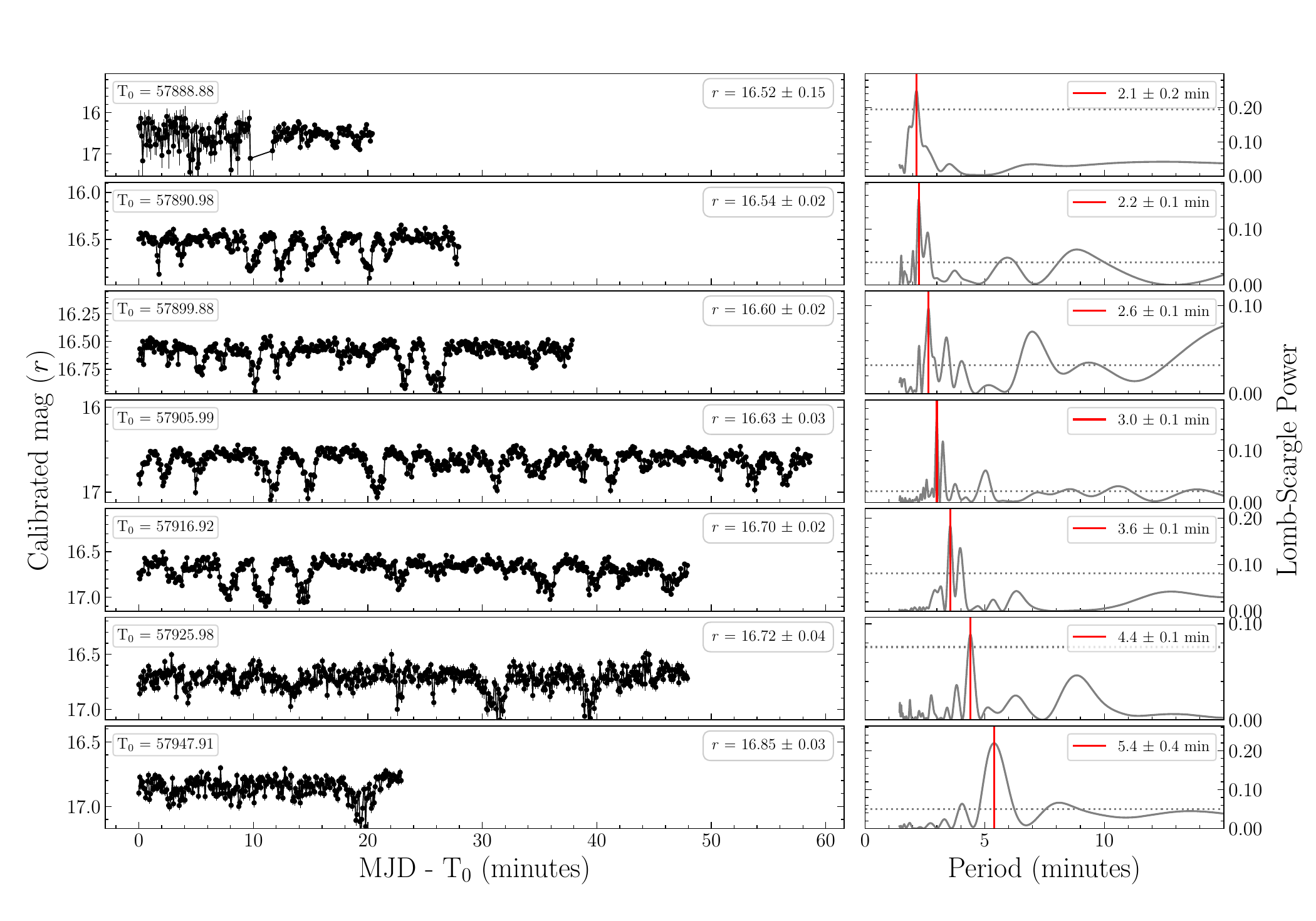}
\caption{From top (epoch \#1) to bottom (\#7), optical light curves for 2017 dataset (left) and the LSP for each epoch (right). The 0.2 FAP and the highest, significant peak are indicated by a horizontal, dashed line and a vertical red line, respectively. }
\label{figA1_2017}
\end{figure*}

We re-analysed the 2017 dataset presented in \citetalias{JimenezIbarra2019b} with the same method used for the 2019 and 2021 datasets (see Sect.~\ref{cap.analysis.peculiarities}). This dataset consisted of seven photometric epochs obtained with the LT between May 15 and July 13, using  the OG515 and KG3 filters ($\sim$\textit{V}~+~\textit{R}). As shown in Fig.~\ref{figA1_2017}, we obtained the same DRPs found by \citetalias{JimenezIbarra2019b}, all of them significant at our 0.2 FAP criterium (i.e. 80\% significance). 

\section{Quiescence optical light curves and periodograms}\label{cap.appendixQuiescence}
\begin{figure*}[!ht]
\centering
\includegraphics[width=\textwidth]{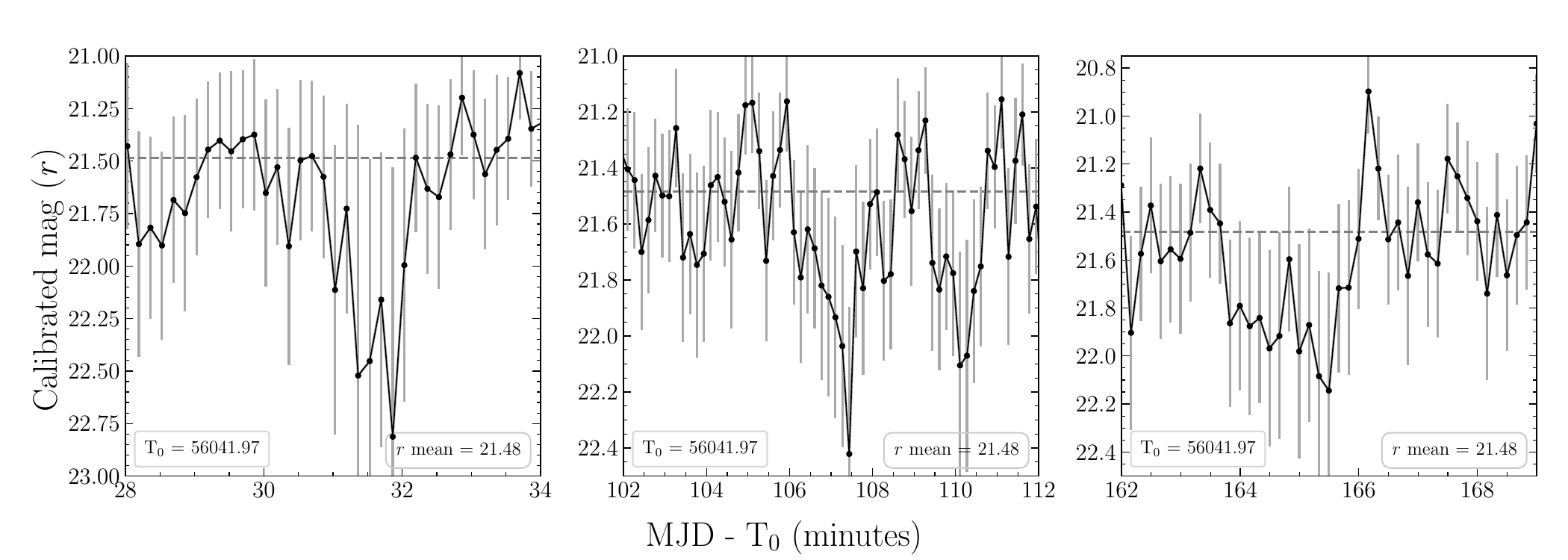}
\caption{Detail of some optical dips observed in epoch \#1 of the 2012 dataset obtained in quiescence.}
\label{figA2_dips}   
\includegraphics[width=\textwidth]{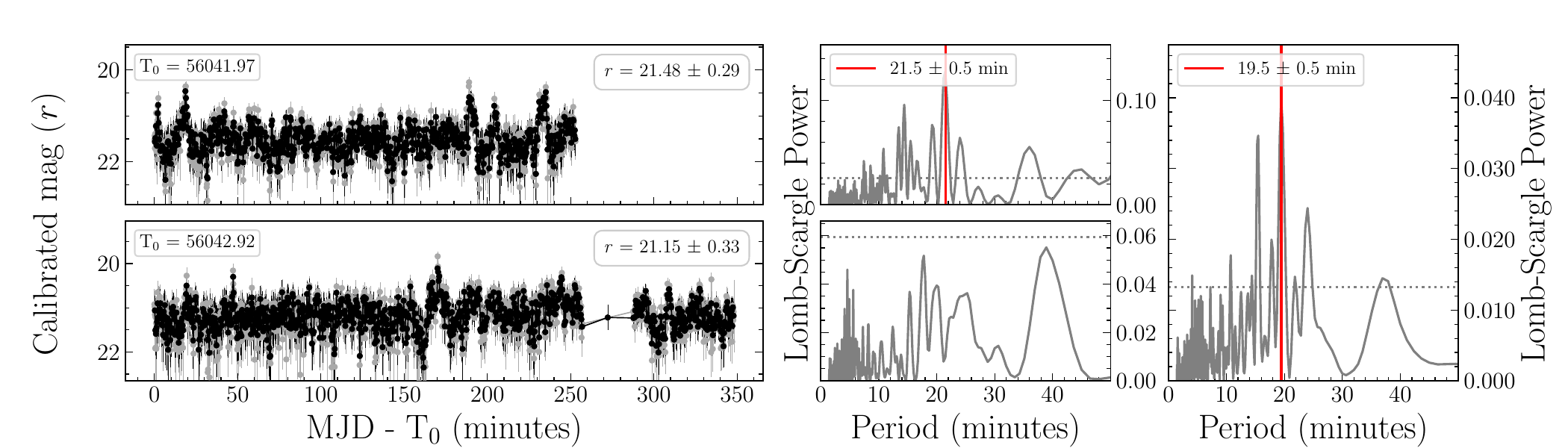}
\caption{Optical \textit{r}-band  
light curves for the 2012 quiescence dataset (left). Original and combined (to 20 s) datasets are plotted in grey and black, respectively. The LSP is computed for the original dataset in each epoch (middle) and for the whole dataset (right). The 0.2 FAP (80\% significance) and the highest, significant peak are indicated by a horizontal, dashed line and a vertical red line, respectively. }
\label{figA2_r_LSPjuntoYSeparado}
\end{figure*}

We analysed quiescent data with the same method used for the 2019 and 2021 datasets (see Sect.~\ref{cap.analysis.peculiarities}). This dataset consisted on two photometric epochs obtained with ULTRACAM \citep{Dhillon2007} at the William Herschel Telescope (Observatorio del Roque de los Muchachos) in two consecutive nights in 2012 (April 24 and 25, with  exposure times of $\sim$10~s and seeings of 1.2 and 1.4 arcsecs, respectively, \citealt{Shahbaz2013}). Optical dips are visible, as we show in Fig.~\ref{figA2_dips}. Both \textit{r}-band light curves and periodograms were analysed (left and middle panels of Fig.~\ref{figA2_r_LSPjuntoYSeparado}), but only the first one revealed a significant peak, with DRP of 21.5$\pm$0.5~min. The second epoch maximum peak is  significant at 79\%. Due to their temporal proximity, we also performed an LSP on the two epochs together (right panel of Fig.~\ref{figA2_r_LSPjuntoYSeparado}), which reveals a significant peak at 19.5$\pm$0.5~min. 

\end{appendix}
\end{document}